\documentclass[
notitlepage
]{revtex4-1}
\usepackage{layout}
\usepackage{graphicx,color}
\usepackage{amsmath,amsfonts,amssymb}
\usepackage{bm}
\usepackage[symbol]{footmisc}
\usepackage{epstopdf}
\usepackage[section]{placeins}
\usepackage{ulem}
\usepackage{enumerate}
\usepackage[final,colorlinks=true,linktocpage=true,urlcolor=red,citecolor=red]{hyperref}

\newcommand{\bcdot}{\bm\cdot}
\newcommand{\bnabla}{\bm\nabla}

\newcommand{\bmu}{\bm u}
\newcommand{\bmQ}{\mathrm{\mathbf{Q}}}
\newcommand{\bmC}{\mathrm{\mathbf{C}}}
\newcommand{\bmS}{\mathrm{\mathbf{S}}}
\newcommand{\bmH}{\mathrm{\mathbf{H}}}
\newcommand{\bmD}{\mathrm{\mathbf{D}}}
\newcommand{\bmI}{\mathrm{\mathbf{I}}}
\newcommand{\trace}[1]{\operatorname{Tr}\left[#1\right]}

\newcommand{\ad}[1]{{\color{blue}#1}}

\begin{document}
\title{Activity pulses induce spontaneous flow reversals in viscoelastic environments}
\author{Emmanuel L. C. VI M. Plan}
\affiliation{Institute of Theoretical and Applied Research, Duy Tan University, Ha Noi 100 000, Viet Nam}
\affiliation{Faculty of Natural Science, Duy Tan University, Da Nang 550 000, Viet Nam}

\author{Julia M. Yeomans}
\affiliation{The Rudolf Peierls Centre for Theoretical Physics, Department of Physics, University of Oxford,\\ Clarendon Laboratory, Oxford OX1 3PU, United Kingdom}

\author{Amin Doostmohammadi}
\affiliation{The Niels Bohr Institute, University of Copenhagen, Blegdamsvej 17, 2100 Copenhagen, Denmark}

%\date{}
\date{\today}

\begin{abstract}
Complex interactions between cellular systems and their surrounding extracellular matrices are emerging as important mechanical regulators of cell functions such as proliferation, motility, and cell death, and such cellular systems are often characterized by pulsating acto-myosin activities. Here, using an active gel model, we numerically explore the spontaneous flow generation by activity pulses in the presence of a viscoelastic medium. The results show that cross-talk between the activity-induced deformations of the viscoelastic surroundings with the time-dependent response of the active medium to these deformations can lead to the reversal of spontaneously generated active flows. We explain the mechanism behind this phenomenon based on the interaction between the active flow and the viscoelastic medium. We show the importance of relaxation timescales of both the polymers and the active particles and provide a phase-space over which such spontaneous flow reversals can be observed. 
Our results suggest new experiments investigating the role of controlled pulses of activity in living systems ensnared in complex mircoenvironments.
\end{abstract}

\maketitle

\section{Introduction}
The study of biological systems as active materials has made tremendous advances in the past decades~\cite{Marchetti13,Bechinger2016,needleman2017active,julicher2018hydrodynamic,doostmohammadi2018active}. The `activity' describes the ability of living systems to extract chemical energy from their surrounding environment and convert it into mechanical work. 
This happens at the level of individual constituents of the matter in, for example, sperm cells thrusting forwards by the rotation of their flagella, bacterial self-propulsion, eukaryotic cells migrating within extracellular networks, and the cytoskeletal machinery inside cells that is powered by motor proteins. 
As such, the overarching theme in various kinds of living systems is the local activity generation that drives the entire system far from thermodynamic equilibrium, resulting in the collective patterns of motion observed in cellular tissues, bacterial colonies, and sub-cellular flows~\cite{ladoux2017mechanobiology,elgeti2015physics,needleman2017active}.

The cross-talk between the mechanical micro-environment of living matter and this intrinsic ability of living systems to actively generate self-sustained motion governs pattern formation and self-organization in important biological processes including collective transport of sperm cells in confined tubes~\cite{hook2020collective}, shaping bacterial biofilms~\cite{vidakovic2018dynamic,meacock2020bacteria}, tissue regeneration~\cite{friedl2009collective}, and sculpting organ development~\cite{maroudas2020topological}. 
Not only does the mechanical micro-environment provide geometrical constraints for active materials~\cite{xi2019material}, it is also often endowed with viscoelastic properties that allow for time-dependent responses to activity-induced stresses and deformations~\cite{Bechinger2016,marchettiarxiv}. 
Significant examples are the extracellular matrices, surrounding cells and tissues that play a key role in cell death and proliferation, stem cell differentiation, cancer migration, and cell response to drugs~\cite{chaudhuri2020effects}. It is thus essential to understand the dynamic interconnection between the activity-induced stresses and the mechanical response of the viscoelastic medium.

Indeed, several recent studies have taken first attempts in this direction, showing that accounting for viscoelastic effects of the medium around living matter results in significant modification in the patterns of motion generated by continuous activity-induced stresses~\cite{BU14,PGGA15,LA16,ZY18,NC18,NBG18,emmanuel2020active}. Importantly, however, in various biological contexts the activity generation is not constant and continuous, but is rather characterized by changes in the activity level and even activity pulses. 
Striking examples are the well-documented acto-myosin contractility pulses that power the activity of epithelial cells and have been shown to be essential in tissue elongation during development~\cite{gorfinkiel2016actomyosin,he2010tissue,martin2009pulsed}. Therefore, here we examine the impact of activity pulses on the behavior of active matter surrounded by a viscoelastic medium.

In order to investigate the fundamental impact of activity pulses we employ a continuum model of active matter based on the theory of active gels, which has proven very successful in describing several aspects of the physics of active systems including acto-myosin dynamics at the cell cortex~\cite{naganathan14,julicher2018hydrodynamic}, acto-myosin induced cell motility~\cite{tjhung17,banerjee20}, actin retrograde flows~\cite{julicher07,Prost15}, and the topological characteristics of actin filaments~\cite{zhang18,kumar2018tunable}. One important prediction of active gel models is the emergence of spontaneous flow generation in a confined active gel~\cite{voituriez2005spontaneous}, which has been further experimentally validated in different biological systems~\cite{duclos2018spontaneous,hardouin19} and is a generic feature of confined active materials. 
Here, we consider a simplified setup of an active gel confined within  viscoelastic surroundings and study the emergence of a spontaneous flow by introducing activity pulses and varying the relaxation time of the viscoelastic medium. We show that introducing activity pulses can result in the reversal of the spontaneous flow direction accompanied by the rearrangement of the orientation of active constituents. The mechanistic basis for this reversal is explained based on the feedback between the active flows and the viscoelastic deformation, particularly in between activity pulses. We further provide a simple model that reproduces the essential dynamics of the flow reversal and shows its dependence on the relevant time scales through a stability-diagram.

The paper is organized as follows. In section~\ref{sec:model} we describe the details of the simulation setup and introduce the governing equations of motion for the active gel, the surrounding viscoelastic medium, and the coupling between the two. The results of the simulation together with the physical mechanism of flow reversal and its associated phase-space are presented in section~\ref{sec:results}. Finally concluding remarks and a discussion of the broader impacts of the results are provided in section~\ref{sec:conc}.

\section{Methods\label{sec:model}}
The active gel is simulated in two dimensions as a horizontal stripe within a passive viscoelastic region, and is differentiated via an indicator function $\phi$ whose value is $\phi=1$ in the active region and $\phi=0$ in the passive viscoelastic region. The indicator function $\phi$ is only defined to distinguish between the active and passive region and as such it is fixed, without any dynamical evolution.
Activity and viscoelasticity are incorporated by introducing a generic two-phase model of active matter in viscoelastic domains~\cite{emmanuel2020active,cates} which is summarised below.

\subsection*{Active region}
Following its success in describing the dynamics of the cell cytoskeleton, bacterial colonies and confluent cell layers, we use liquid crystal nematohydrodynamics to model the active region~\cite{ramaswamy2010mechanics,doostmohammadi2016defect,doostmohammadi2018active}.
Within this framework each active particle generates a dipolar flow field with axis along its direction of alignment. The alignment direction is nematic, i.e. it has a head-tail symmetry. This can be captured on a coarse-grained level by the order parameter tensor $\bmQ $  through its principal eigenvector, which describes the nematic director orientation, and the associated eigenvalue, which describes the degree of alignment.

The free energy $f_Q$ for two-phase nematic models follows the Landau-De Gennes description
\begin{equation}
f_{Q}=A_Q\left[\dfrac{1}{2}\left(1-\dfrac{\eta(\phi)}{3}\right)\trace{\bmQ^2}-\dfrac{\eta(\phi)}{3}\trace{\bmQ^3}+\dfrac{\eta(\phi)}{4}\trace{\bmQ^2}^2\right]+\dfrac{K_Q}{2}(\bnabla\bmQ)^2+L_0(\bnabla\phi\bcdot\bmQ\bcdot\bnabla\phi)
\end{equation}
where $A_Q$ describes the stability of the nematic or isotropic configurations, with the former being favoured when $\eta>2.7$. The elastic coefficient $K_Q$ penalizes gradients in $\bmQ$, and a positive (negative) $L_0$ enforces nematic orientation parallel (perpendicular) to the active-viscoelastic interface.

In the presence of a velocity field $\bmu$, the nematic tensor is evolved according to the equation $
\partial_t\bmQ+\bmu\bcdot\bnabla\bmQ 
=\bmS_Q+\Gamma_Q\bmH_Q$
where the left-hand side is the usual material advective derivative. The co-rotational term $\bmS_Q=(\xi\bmD+\bm\Omega)(\bmQ+\bmI/3)+(\bmQ+\bmI/3)(\xi\bmD-\bm\Omega)-2\xi(\bmQ+\bmI/3)\trace{\bmQ\bnabla\bmu}$ describes nematic reorientation in response to both vorticity $\bm\Omega$ and flow strain $\bmD$, with the tumbling parameter $\xi$ determining whether the directors align or tumble in the flow. $\Gamma_Q$ controls the speed of relaxation towards the free energy minimum determined by the molecular field $\bmH_Q=-\delta f_Q/\delta \bmQ$. The typical nematic relaxation timescale $t_n$ when confined in a channel of width $W$ is given by $W^2/\Gamma_Q K_Q$ and the dynamical equation for $\bmQ$ can thus be rewritten as 
\begin{equation}
\partial_t\bmQ+\bmu\bcdot\bnabla\bmQ 
=\bmS_Q+(W^2/t_nK_Q)\bmH_Q.
\label{eq:Q}
\end{equation}

Since individual components of the active region generate dipolar forces with axis along their direction of alignment, the corresponding active stress is proportional to the orientation tensor~\cite{ramaswamy2010mechanics,Marchetti13,Prost15}
\begin{equation}
\bm\sigma_{\rm active}=-\zeta\phi\bmQ
\label{eq:actstress}
\end{equation}
such that gradients in the orientation field generate forces on the fluid and drive active flows. The activity parameter, $\zeta$, measures the strength of the active driving.

\subsection*{Viscoelastic region}
The passive region is endowed with viscoelasticity that is described by the conformation tensor $\bmC$, which characterizes the polymer orientation by its principal eigenvector, and the (square of the) polymer length, by its trace. 
Here we use the Oldroyd-B model to reproduce simple viscoelastic effects, i.e. polymer relaxation linear with respect to the elongation, and governed by a single relaxation time $\tau$~\cite{BC18}. 

The free energy associated with an Oldroyd fluid 
\begin{equation}
f_{C}=A_C(1-\phi)(\trace{\bmC-\bmI}-\ln\operatorname{det}\bmC)/2,
\end{equation} 
governs the polymer relaxation to its unstretched equilibrium $\bmC=\bmI$ (where $\bmI$ is the identity tensor). Here, the modulus of elasticity $A_C=\nu/\tau$ is the ratio of the polymer contribution to viscosity $\nu$ to the polymer relaxation time $\tau$. The corresponding molecular field $\bmH_C=-\delta f_C/\delta\bmC$, appears in the dynamical equation governing the evolution of $\bmC$: $\partial_t\bmC+\bmu\bcdot\bnabla\bmC =\bmS_C+\Gamma_C[\bmH_C\bmC+\bmC^\top\bmH_C^\top]$, where $^\top$ denotes the matrix transpose. Similarly to Eq.~\eqref{eq:Q} for the orientation tensor $\bmQ$, the evolution of $\bmC$ accounts for the advection of the polymer conformation $\bmC$, its response to velocity gradients through $\bmS_C=\bmC\bm\Omega-\bm\Omega\bmC+\bmC\bmD+\bmD^\top\bmC^\top$, and its relaxation to equilibrium at a rate $\Gamma_C=\nu^{-1}$. The viscoelastic timescale is clearly seen when this equation of motion is simplified as 
\begin{equation}
\partial_t\bmC+\bmu\bcdot\bnabla\bmC =\bmS_C-\cfrac{1}{\tau}(1-\phi)(\bmC-\bmI).
    \label{eq:C}
\end{equation}
The polymer contribution to the stress, within the assumptions of the Oldroyd-B model, is
\begin{equation}
\bm\sigma_{\rm polymer}=A_C(1-\phi)(\bmC-\bmI),
\label{eq:polystress}
\end{equation} 
where the factor $1-\phi$ ensures that the polymer stress only acts within the passive region (where $\phi=0$).

\subsection*{Coupling and simulation details}
The active and viscoelastic regions interact with each other through the velocity field $\bmu$ which obeys the incompressible Navier-Stokes equations,
\begin{equation}
    \rho (\partial_t \bmu + \bmu \bcdot \bnabla \bmu)=-\bnabla p+\bnabla\bcdot\bm\sigma, \quad(\bnabla\bm\cdot\bmu=\bm0),
\label{eq:NSE}
\end{equation}
where $\rho$ is the fluid density, $p$ is the pressure, and $\bm\sigma$ is the sum of viscous, capillary, elastic, active (Eq.~\eqref{eq:actstress}), and polymer (Eq.~\eqref{eq:polystress}) stresses. Both the active and the viscoelastic regions exert stresses on the fluid, and the resulting velocity field couples the two regions through advective and co-rotational terms in Eqs.~\eqref{eq:Q} and \eqref{eq:C}. The explicit forms of the other stresses are given in~\cite{emmanuel2020active}.

\begin{figure}[ht]
\includegraphics[width=0.6\textwidth]{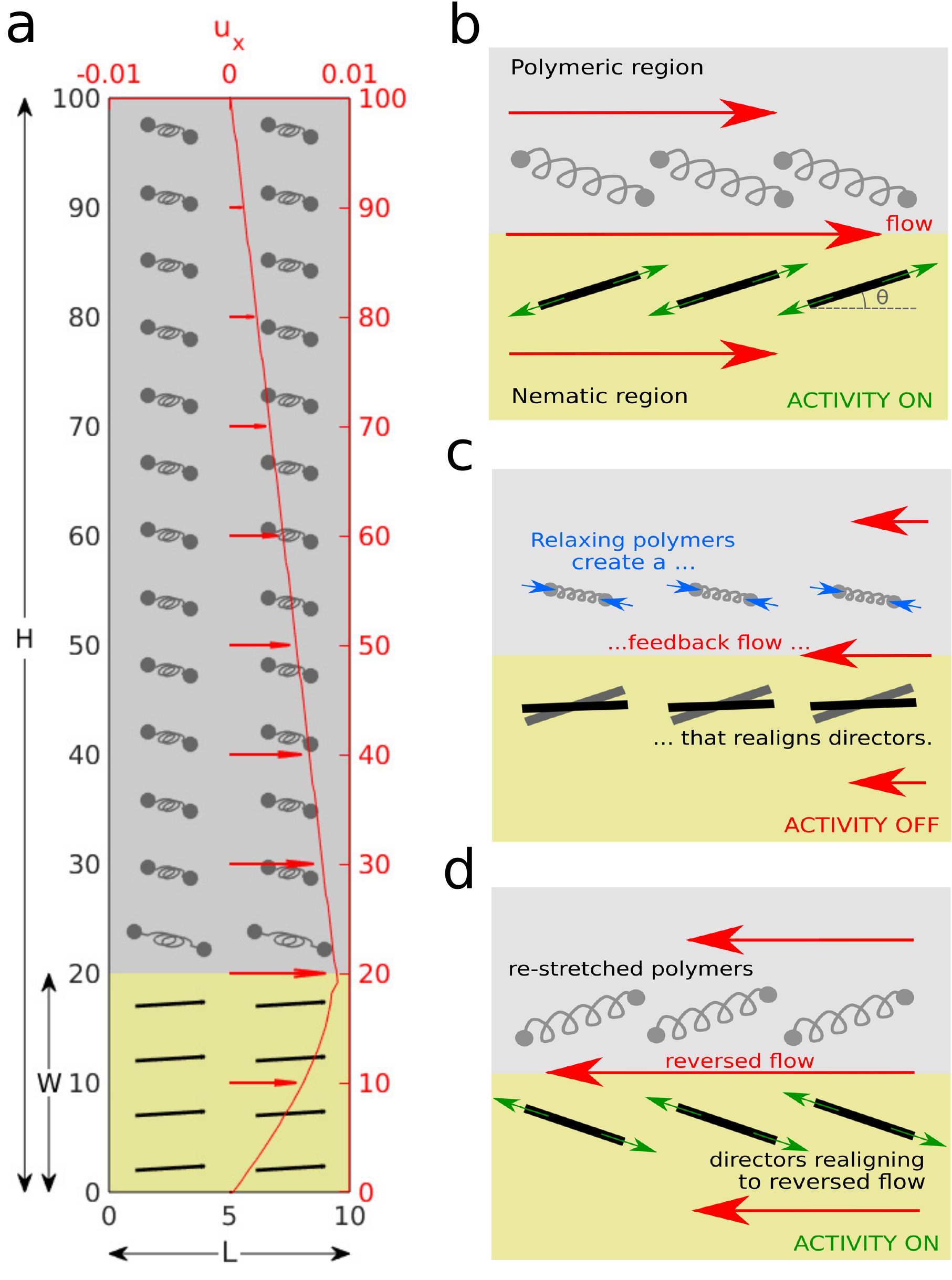}
\caption{{\bf Schematic of problem set up and flow reversal mechanism.} (a) The simulation domain indicating the active region (gold) with black nematic directors and the viscoelastic region (silver) with polymers in grey. A steady shear-like flow in both regions is shown in red ($\zeta=0.011,\tau=5000,t_n=4000$). (b)-(d) The mechanism behind the flow reversal: (b) The flow stretches the polymers and orients both polymers and nematic directors ($\theta_n$) to the Leslie angle; (c) When activity is turned off, the polymers relax and create a weaker but reversed flow, which reorients the directors; (d) When activity resumes, the directors either return to their original steady state or reverse direction depending on the balance between their orientation and the residual polymer stress.}
\label{fig:schematic}
\end{figure}

Equations \eqref{eq:Q},\eqref{eq:C}, and \eqref{eq:NSE} are evolved using a hybrid lattice Boltzmann method~\ad{\cite{Marenduzzo2007,thampi2014instabilities}}.
The simulation domain has dimensions $L\times H$, is periodic in the $x$-direction, and has no-slip boundary conditions in the $y$-direction (see schematic in Fig.~\ref{fig:schematic}(a)). 
The  dynamics is not affected by the length of the channel $L$ because of periodicity; our results here use $L=10,H=100$.
The width of the active region is fixed at $W=20$.

%\ju{REORDERED}\\
In order to obtain a unidirectional flow the parameters for the active region are chosen to be $A_Q=1.0$ and $K_Q=0.2$, and we use $\xi=0.7$ corresponding to the flow aligning regime. $\eta(\phi)$ is chosen such that $\eta=2.95$ within the active region $\phi=1$~\cite{chandragiri2019active}, while the passive viscoelastic region, $\phi=0$, is in the isotropic phase with $\eta<2.7$. The polymer contribution to viscosity is fixed at $\nu=1$ and $A_C=1/\tau$. The value of the flow parameters are  $\rho=1,~p=0.25$, and the Newtonian contribution to the viscosity is $\nu_{\rm flow}=2/3$. 
We also enforce weak nematic anchoring at the boundary, $L_0=0.05$, for stability. This was verified to have no qualitative effect on the flow reversal dynamics.

Each simulation begins with equilibriated polymers $\bmC=\bmI$ and nematic directors with small random perturbation about the $x$-axis. Active stresses are applied until the system establishes a steady-state, unidirectional flow, after which the activity is temporarily turned off at $t_{\rm off}=10^5$ for a duration of $d=t_{\rm on}-t_{\rm off}$ time steps. The equations are solved until the flow is re-established.

\section{Results\label{sec:results}}
When activity is turned off at a time $t_{\rm off}$, and turned back on at a time $t_{\rm on}$, the flow is re-established in the same or, surprisingly, in the opposite direction. This is not a random choice but depends sensitively on the parameters setting the relevant nematic and viscoelastic time scales. For example, the supplementary Movie shows several successive reversals in the direction of the velocity (see Data Availability). %\Em{[\url{https://drive.google.com/file/d/1BcMZCo9TYWDhPAbDKVFyliRrb6jeRXoS/view?usp=sharing}][Note: Upload in youtube or label accordingly as Supplementary.]}
We first explain how this dependence comes about, and then present a simple model which illustrates the underlying physics. The mechanism of the flow reversal is summarised in Fig.~\ref{fig:schematic}(b)--(d).

\begin{figure}[ht]
    \centering
    \includegraphics[width=0.99\textwidth]{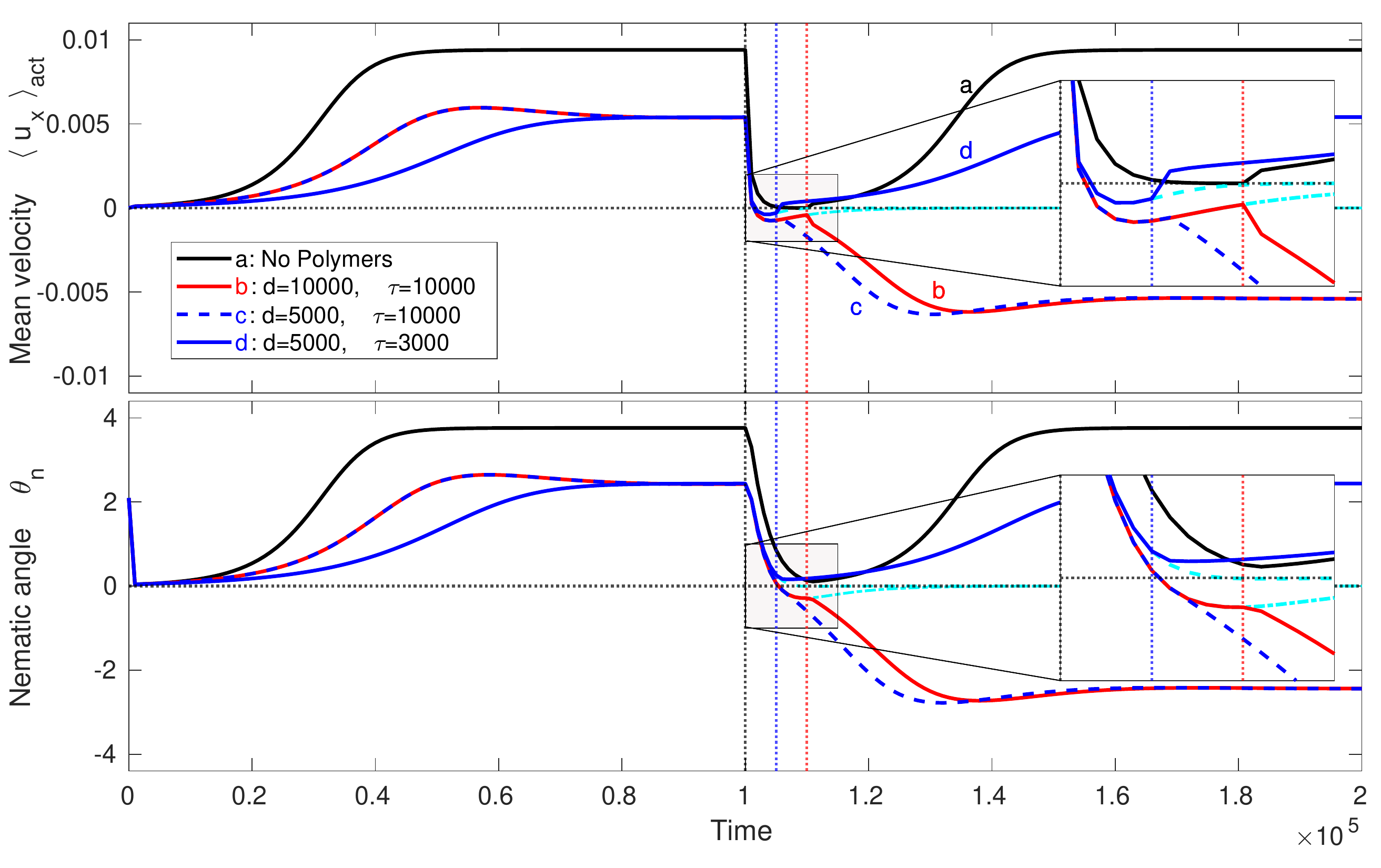}
    \includegraphics[width=0.99\textwidth]{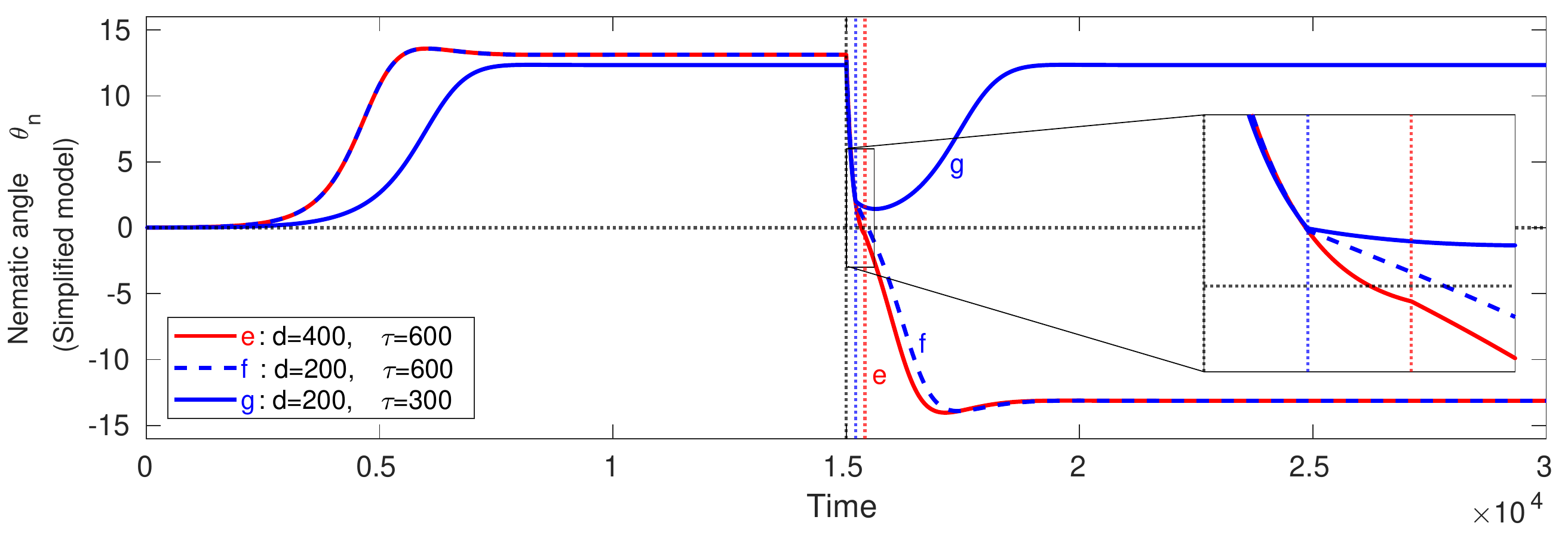}
    \caption{{\bf Emergence of spontaneous flow reversals.} (Top) The mean velocity $\langle u_x\rangle_{\rm act}$ in the active stripe and (middle) the nematic angle at the interface $\theta_n$ as a function of time for the full simulations ($t_n=4000$), and (bottom) the nematic angle for the simplified model ($t_n=150$). (a) black: no polymers, activity switching time $d=10^4$; (b) red: activity switching time $d=10^4$, polymer relaxation time $\tau=10^4$; (c) dashed: activity switching time $d=5000$, polymer relaxation time $\tau=10^4$; (d) blue: activity switching time $d=5000$, polymer relaxation time $\tau=3000$. The cyan curves show the behaviour in the absence of reactivation. (e) red: activity switching time $d=400$, polymer relaxation time $\tau=600$; (f) dashed: activity switching time $d=200$, polymer relaxation time $\tau=600$; (g) blue: activity switching time $d=200$, polymer relaxation time $\tau=300$. The insets show a zoom in on the period of inactivity and immediately after reactivation.}
    \label{fig:cycle}
\end{figure}

Fig.~\ref{fig:cycle} shows the variation of the mean velocity in the channel, $\langle u_x\rangle$, and the angle $\theta_n$ that the mean director field at the interface makes with the channel axis as a function of time for selected simulation parameters. 
Consider first Fig.~\ref{fig:cycle}(a) where there are no polymers in the passive region. Activity is switched on at time $t=0$. This drives the active nematic instability and active stresses resulting from gradients in the director field set up a linear flow along the stripe.
The spontaneous flow is stabilised by the channel interfaces, and the resulting steady-state flow profile corresponds to directors aligning at the Leslie angle to the local shear~\cite{DeGennesBook,thijssen2020binding}. 
When the activity is switched off the flow decays faster than the relaxation time of the nematic director. 
In the absence of polymers if activity is switched back on before the end of the decay the residual director rotation ensures that the flow re-starts in the same direction as before.

The presence of polymers changes the response of flow to activity pulses. In Fig.~\ref{fig:cycle}(b)--(d) the passive medium is viscoelastic and the shear flow induced by the activity stretches the polymers thus storing energy. The initial build-up of the flow is similar to the no polymer case, but slower, as work is done to stretch the polymers, and the steady-state value of the flow in the channel is lower because the polymers impose stresses that oppose the flow. This also means that after $t_{\rm off}$ the decays of the velocity and director fields to equilibrium are faster and, in particular, stress imposed by the relaxation of the surrounding polymers may be strong enough to reverse the flow in the channel.

For example in Fig.~\ref{fig:cycle}(b) both the residual flow and the director have reversed at $t_{\rm on}$ making a velocity reversal inevitable. 
By comparison, for the same polymer relaxation time but faster switching of activity in  Fig.~\ref{fig:cycle}(c), at  $t_{\rm on}$ the director distortion is still just positive but the velocity has been reversed by the polymer stresses. The activity tries to rebuild the flow in the original direction whereas the remaining elastic energy stored in the surrounding polymers pushes the fluid in the opposite direction. The flow slowly reverses as the instability is (just) overcome by the residual polymer stresses. The director distortion reverses and the velocity slowly increases until it eventually attains its steady-state value, but in the opposite direction.  
By contrast Fig.~\ref{fig:cycle}(d) shows an example of faster polymer relaxation, compared to Fig.~\ref{fig:cycle}(c), where the polymer stress is not sufficiently strong to cause flow re-orientation and both nematic directors and mean velocity regain their original direction.

Such a flow reversal mechanism that depends on stored polymer stresses and residual director orientation implies that the relevant timescales here are the polymer relaxation time $\tau$, the nematic relaxation time $t_n$ and the period of inactivity $d=t_{\rm on}-t_{\rm off}$. To further explain the phenomenon of flow reversal and highlight the competition between these varying time scales, we construct simplified, space-independent, dynamic equations for the evolution of the active nematic and polymeric particles.

To this end, we consider the dynamics of the alignment for the angles $\theta_n,\theta_p$ formed by the active nematic directors and the polymer directors, respectively, with respect to the direction of a simple shear flow, $\bmu=(\dot{\gamma}y,0)$~\cite{aigouy2010cell}:
\begin{eqnarray}
\dfrac{d\theta_n}{dt}&=&\dot{\gamma}\xi_n\cos 2\theta_n-\dfrac{1}{t_n}\theta_n,\label{eq:poor1}\\
\dfrac{d\theta_p}{dt}&=&\dot{\gamma}\xi_p\cos 2\theta_p-\dfrac{1}{\tau}\theta_p.
\label{eq:poor2}
\end{eqnarray}
These equations are approximations of the orientation of rod-like particles with tumbling parameters $\xi_n,\xi_p$ in response to a shear flow with rate $\dot{\gamma}$. In the absence of shear, $\dot{\gamma}=0$, the angles will relax exponentially to zero.

A time-dependent shear rate $\dot{\gamma}$ couples the two equations and can be determined from balancing the viscous stress $\bm\sigma_{\rm viscous}=2\nu_{\rm flow}\bmD$, with the active (Eq.~\eqref{eq:actstress}), and polymer (Eq.~\eqref{eq:polystress}) stresses. Solving for the off-diagonal term of the rate of strain tensor $\bmD$ and constructing the tensors $Q_{ij}=n_in_j-\delta_{ij}/2$ and $C_{ij}=p_ip_j$ from the unit directors $\bm n=(\cos\theta_n,\sin\theta_n)$ (and analogously for $\bm p$), we obtain the simple time-dependent shear
\begin{equation}
\dot{\gamma}=D_{xy}=\dfrac{1}{2\nu_{\rm flow}}\left(\zeta Q_{xy}-\frac{\nu}{\tau} C_{xy}\right)=\dfrac{1}{4\nu_{\rm flow}}\left(\zeta \sin 2\theta_n-\frac{\nu}{\tau}\sin 2\theta_p\right),
\label{eq:shearrate}
\end{equation}
in terms of the nematic and polymer directors. For simplicity, we fix $\nu_{\rm flow}=2/3,\nu=1$ and set $\xi_n=1.1$ and $\xi_p=0.275$ and take initial conditions $\theta_n=\theta_p=0.01$.

It is clear from \eqref{eq:shearrate} why, when the activity $\zeta=0$ is turned off, the flow reverses in the opposite direction due to the presence of the polymers. Moreover, a simulation of this simpler set of equations features the same essential reason for the reversal as in the full simulations: there exists sufficient polymer stress to reorient the nematics during the period between activity pulses. 
Figures~\ref{fig:cycle}(e)--(g) show that the nematic angle $\theta_n$ exhibits trajectories similar to the result from the full equations of motion, Figs.~\ref{fig:cycle}(b)--(d). The reversed flow can drive the nematic angle negative during the period of inactivity, which is a sufficient condition for a flow reversal. Moreover, even if $\theta_n$ remains positive when the activity is switched on, the polymer stress can still overcome the activity (i.e. $\zeta \sin 2\theta_n<\nu\sin 2\theta_p/\tau$) and continue to drive the nematic director to reverse direction.

\begin{figure}
    \centering
    \includegraphics[width=0.49\textwidth]{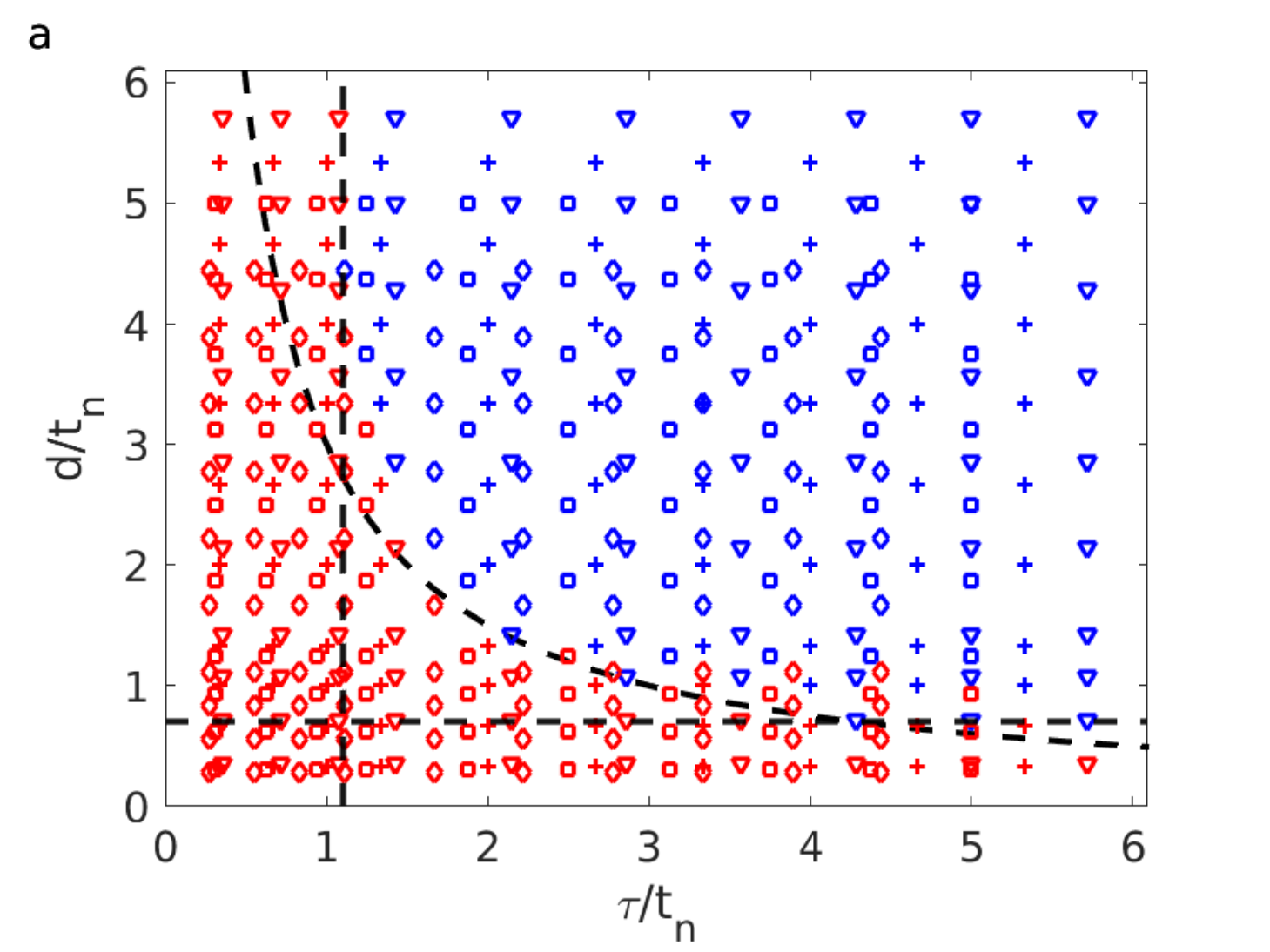}
    \includegraphics[width=0.49\textwidth]{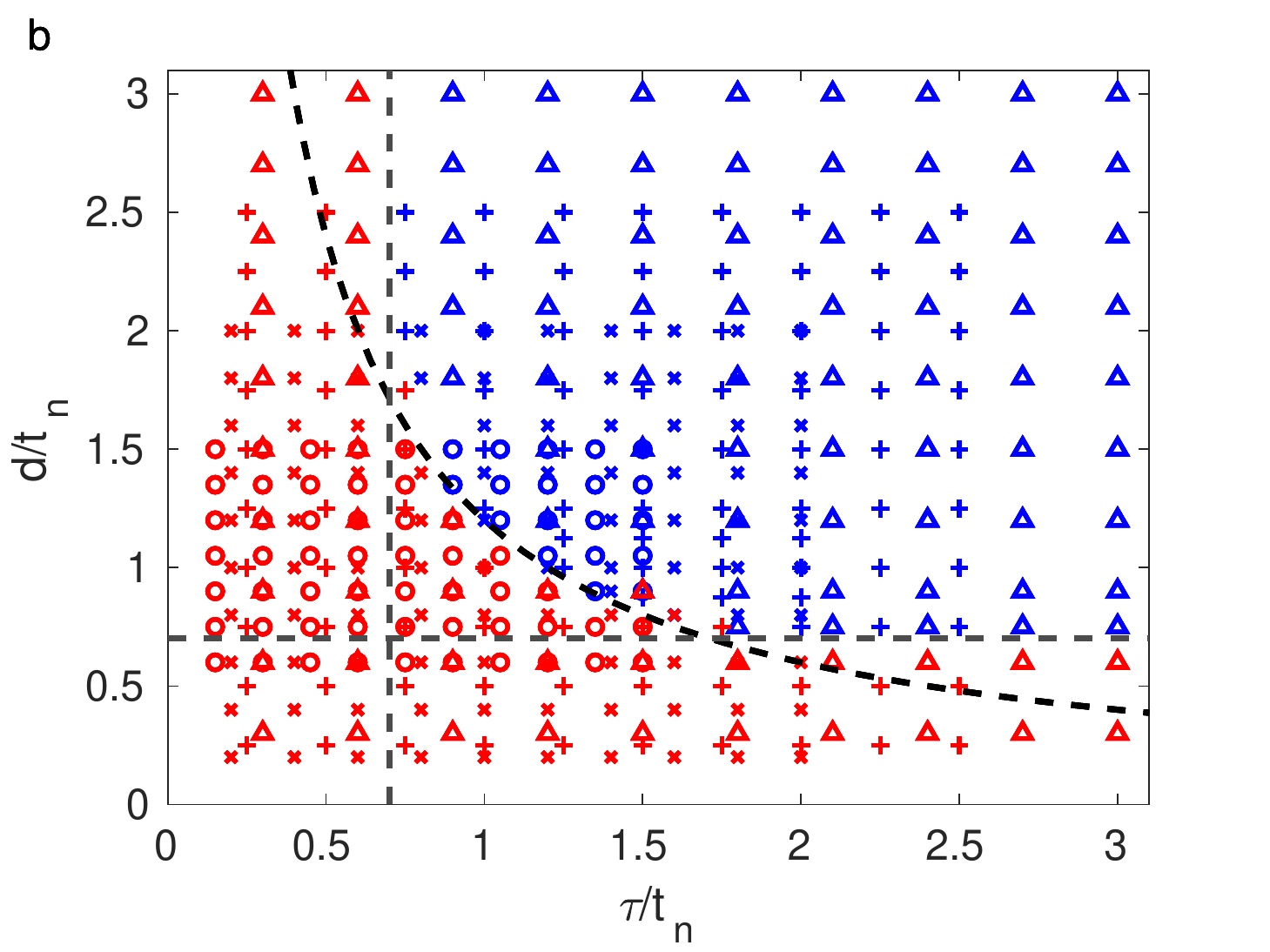}
    \caption{\textbf{Phase diagram of the (a) simplified equations and (b) full simulations.} Blue points indicate flow reversal while red points indicate non-reversal. Different markers represent different nematic relaxation timescales: 
    (a) $t_n=140$ ($\bigtriangledown$); $t_n=150$ ($+$); $t_n=160$ ($\square$); $t_n=180$ ($\diamond$); (b) $t_n\approx6667$ ($\bigtriangleup$); $t_n=5000$ ($+$); $t_n=4000$ ($\times$); $t_n\approx3333$ ($\circ$). The dashed lines are (a) $\tau/t_n>1.1, d/t_n>0.7, d\tau /t_n^2>3$ and (b) $\tau/t_n>0.7, d/t_n>0.7,  d\tau/t_n^2>1.2$.
    }
    \label{fig:phased}
\end{figure}

Using the simple model we examined the phase space of flow reversal for varying time scales of nematic and polymer relaxations, $t_n$ and $\tau$ respectively, as well as the activity switching time $d=t_{\text{on}}-t_{\text{off}}$ (Fig.~\ref{fig:phased}(a)).
In particular, we observe that the reversal can be ensured if three dimensionless ratios are sufficiently large:
\begin{enumerate}[(i)]
\item $d/t_n$: nematic directors have enough time to relax
\item $\tau/t_n$:  polymers retain enough energy to reverse nematic orientation when $\theta_n\approx 0$ 
\item $d\tau/t_n^2$: residual polymer stress upon moment of reactivation can help reverse nematic orientation if condition (i) alone is insufficient
\end{enumerate}
These three constraints are illustrated as dashed black lines on the flow reversal phase diagram in the $d/t_n$-$\tau/t_n$ phase space and clearly distinguish the parameter space with ({\it blue markers}) and without ({\it red markers}) flow reversal. 

We also plot, in Fig.~\ref{fig:phased}(b), the similar phase diagram obtained from full simulations of the active gel surrounded by a viscoelastic medium showing close qualitative agreement with the simple model, and indicating that the parameter region for flow reversal is suitably captured by these three constraints.
The quantitative difference between the simplified model and the full simulations is expected because in the simplified model there is no space dependence and we only account for polymer orientation in the polymer stress, whereas the full simulations are space-dependent and have a polymer stress (Eq.~\eqref{eq:polystress}) that depends on polymer elongation, which is higher when $\tau$ is larger.
Notwithstanding these limitations, the close qualitative agreement of the angle profiles and the phase diagram obtained from the simple model with those from the full hydrodynamic simulations of spatio-temporal evolution of the active nematics and polymeric fluids
confirms the importance of the time scale constraints for the flow reversal and the underlying physics of stress balance during the period of inactivity.

\section{Conclusion\label{sec:conc}}

In this report we present how, in the absence of any external forcing, activity pulses in living matter interacting with a viscoelastic environment can spontaneously generate flow reversals. Based on a well-documented active gel theory, a spontaneous steady flow of active matter is achieved even while in contact with polymer-laden surrounding. This flow stretches the polymers near the interface which, in between periods of activity, relax and produce a weak backflow that may determine the flow direction upon resumption of the active driving.

The well-established spontaneous flow generation of confined active matter relies on the level of activity. Here the spontaneous flow reversals hinge on several timescales: the polymer relaxation time, the interval between activity pulses, and the relaxation dynamics of nematic active matter, as shown in the phase diagram. Indeed the need for sufficient polymer stretching and feedback as well as quick nematic reordering highlight the time-dependent viscoelastic response in between activity pulses. Our work emphasizes not only the importance of accounting for a viscoelastic environment, but also the involvement of several timescales arising from both active matter and its surroundings.

Our theoretical work invites several experiments to be performed in order to confirm our predictions and deepen our understanding of the role of a viscoelastic environment on the dynamics of living matter. For instance, cells confined in a monolayer have been shown to display nematic behaviour and exhibit unidirectional flows in channels \cite{STEDDEN2019908}. 
They could be repeatedly subjected to cell inhibitors or uncouplers \cite{jacobelli,kolega,D0SM00944J} to simulate gaps between activity pulses. 
Another option would be to regulate the movement of a colony of elongated bacteria by utilizing phototactic methods \cite{wilde}. 
Our result thus holds potential in understanding mechanotaxis and motivating the use of viscoelastic media to control living matter at microscopic scales.

\section*{Acknowledgement and Funding}
We thank Sally Horne-Badovinac for helpful discussions. A.D. acknowledges support from the Novo Nordisk Foundation (grant no. NNF18SA0035142), Villum Fonden (grant no. 29476), Danish Council for Independent Research, Natural Sciences (DFF-117155-1001), and funding from the European Union’s Horizon 2020 research and innovation program under the Marie Sk\l odowska-Curie grant agreement no. 847523 (INTERACTIONS).

\section*{Data Availability}
The codes used for this work are available in \url{https://github.com/elcplan/Activity_pulses_flow_reversal-Plan_Doostmohammadi_Yeomans.git}. The supplementary movie (in .gif and .mp4 format) is also available in this repository.

\section*{Competing Interests}
The authors declare no competing interests.

\section*{Authors contribution}
All authors contributed in the conceptualisation, data analysis, and the preparation and submission of the manuscript.

\end{document}